\documentclass[%
 reprint,
 superscriptaddress,
 amsmath,
 amssymb,
 aps,
 pra,
]{revtex4-2}

\usepackage{graphicx}
\usepackage{dcolumn}
\usepackage{bm}
\usepackage{hyperref}

\usepackage{float}
\usepackage{blindtext}
\usepackage{amsmath,amsfonts,amsthm,physics}
\usepackage{comment}                            
\usepackage{lipsum}
\usepackage[export]{adjustbox}
\usepackage{xprintlen}

\usepackage{placeins}

\usepackage{xcolor}
\definecolor{sout}{HTML}{EBEBEB}

\newcommand{\bs}[1]{\boldsymbol{#1}}
\newcommand{\LL}{\left}
\newcommand{\RR}{\right}
\newcommand{\id}{\mathbb{I}}
\newcommand{\sz}{\sigma_Z}

\renewcommand{\selectlanguage}[1]{}
\newcommand{\E}{\mathcal{E}}
\newcommand{\bigO}{\mathcal{O}}
\newcommand{\dee}{\ \mathrm{d}}
\newcommand{\T}[1]{\text{#1}}
\newcommand{\R}{\mathcal{R}}

\begin{document}

\preprint{APS/123-QED}

\title{Entanglement distribution via satellite: an evaluation of competing protocols assuming realistic free-space optical channels}

\author{Nicholas Zaunders} \email{n.zaunders@uq.edu.au}
\affiliation{
    Centre for Quantum Computation and Communication Technology, School of Mathematics and Physics, University of Queensland, St Lucia, Queensland 4072, Australia.
}

\author{Timothy C. Ralph}
\affiliation{
    Centre for Quantum Computation and Communication Technology, School of Mathematics and Physics, University of Queensland, St Lucia, Queensland 4072, Australia.
}

\begin{abstract}
A key technical requirement of any future quantum network is the ability to distribute quantum-entangled resources between two spatially separated points at a high rate and high fidelity. Entanglement distribution protocols based on satellite platforms, which transmit and receive quantum resources directly via free-space optical propagation, are therefore excellent candidates for quantum networking, since the geometry and loss characteristics of satellite networks feasibly allow for up to continental-scale ($\sim10^3$ km) over-the-horizon communication without the infrastructure, cost, or losses associated with equivalent fibre-optic networks. In this work, we explore two network topologies commonly associated with quantum networks - entanglement distribution between two satellites in low-Earth orbit mediated by a third satellite and entanglement distribution between two ground stations mediated by a satellite in low-Earth orbit, and two entanglement distribution schemes - one where the central satellite is used as a relay, and the other where the central satellite is used to generate and distribute the entangled resource directly. We compute a bound on the rate of distribution of distillable entanglement achieved by each protocol in each network topology as a function of the network channels for both single-rail discrete- (DV) and continuous-variable (CV) resources and use or non-use of probabilistic noiseless linear quantum amplification (NLA). In the case of atmospheric channels we take into account the turbulent and optical properties of the free-space propagation. We determine that for the triple-satellite network configuration, the optimal strategy is to perform a distributed NLA scheme in either CV or DV, and for the ground-satellite-ground network the optimal strategy is to distribute a DV resource via the central satellite.
\end{abstract}

\maketitle

\section{Introduction} \label{sec:intro}

Modern quantum protocols leverage the nonclassical properties of quantum states to perform tasks that are otherwise impossible using classical technology \cite{nielsen_quantum_2010}, including quantum key distribution \cite{bennett_quantum_2014, ekert_quantum_1991, gisin_quantum_2002, ralph_continuous_1999, ralph_security_2000}, teleportation of quantum states \cite{bennett_teleporting_1993, bennett_purification_1996, braunstein_criteria_2000}, clock synchronisation \cite{jozsa_quantum_2000, yurtsever_lorentz-invariant_2002, komar_quantum_2014, ilo-okeke_remote_2018}, distributed computing \cite{cirac_distributed_1999} and quantum sensing \cite{gottesman_longer-baseline_2012, degen_quantum_2017, zhuang_distributed_2018, xia_repeater-enhanced_2019}. In quantum protocols, advantage is commonly obtained by leveraging quantum resources with high degrees of entanglement \cite{azuma_quantum_2023}, and the degree of advantage for many protocols can be described in terms of the amount of entanglement shared between the two parties who wish to communicate some classical or quantum information (e.g. \cite{ekert_quantum_1991, shor_simple_2000}). To this end, the efficient and high-volume distribution of entangled resources between two parties who may be separated by some large distance is a key focus of quantum communication in general, and new developments in the field of entanglement distribution translate readily to improvements in many other areas of quantum communication \cite{azuma_quantum_2023}.

Further to this is the notion of satellite-based quantum communication \textcolor{black}{\cite{oi_cubesat_2017, bedington_progress_2017, simon_towards_2017, khatri_spooky_2021, gonzalez-raya_satellite-based_2024, pirandola_satellite_2021, ghalaii_quantum_2022}}, which has been driven by recent development into classical free-space-optical satellite networks. In the standard fibre-optic links used for classical optical-frequency telecommunications, the transmissivity $\eta$ of the quantum channel decays exponentially with distance \cite{zhang_long-distance_2020}. Theoretical bounds on the capacity of quantum channels limit the maximum rate of distribution of an entangled resource to $-\log(1-\eta)$ \cite{pirandola_fundamental_2017}, which suggest that the effectiveness of fibre-based entanglement distribution must decay exponentially fast with distance as well. This limitation can be circumvented by instead broadcasting part or all of the optical entangled state via a free-space atmospheric channel, which differ from fibre-optic channels in that they have stochastic loss profiles arising from atmospheric effects such as turbulence \cite{schmidt_numerical_2010, andrews_laser_2023, vasylyev_toward_2012, vasylyev_atmospheric_2016}. In a configuration where one or more satellite relays are connecting two ground stations, the near-constant loss profile of free-space channels allow satellite-based quantum communication to deliver effective communication even for nominal distances on a continental scale ($10^3$ km) \cite{yin_satellite-based_2017, yin_entanglement-based_2020, ren_ground--satellite_2017, liao_satellite--ground_2017}. Such a feat would be essentially impossible on an equivalent fibre link without a large number of relay or repeater stations installed along the line. 

In this work, we provide a comparative study on two protocols which utilise a station-relay-station configuration to achieve the same goal (distribution of an entangled source between the two distant stations manned by Alice and Bob) in different ways. In the first scheme, which we call the `distribution' configuration, the entangled resource is generated first at the central relay, manned by Charlie, and each arm of the state is sent directly to the distant stations, manned by Alice and Bob respectively. In the second scheme, which we call the `relay' configuration, the entangled resource is generated first at Alice's station, and one arm of the resource is sent to the relay. At the same time, an auxiliary entangled mode is generated by Bob's station and sent via a separate channel to Charlie. Charlie then measures the incoming modes and performs entanglement swapping \cite{zukowski_event-ready-detectors_1993}, leaving Alice and Bob with a mutually entangled state. \textcolor{black}{Note that the relay scheme, based on the teleamplification protocols described in \cite{mauron_comparison_2022} and \cite{winnel_overcoming_2021} for DV and CV respectively, is distinct from MDI QKD schemes such as the one described in \cite{cao_long-distance_2020} despite having similar network architecture. Whilst Charlie is an untrusted node (as in MDI), his main purpose is so Bob achieves a performance advantage (not seen in MDI protocols) by manipulating his ancilla entangled state.} We investigate these protocols for both a network topology involving three satellites, each in low-Earth orbit, communicating via symmetric diffraction-limited optical channels as well as a topology in which two ground stations communicate through atmospheric channels to a central relay satellite.

It is known that in free-space optical satellite communication with ground stations, uplink channels present a significantly higher and more variable loss profile as a result of turbulent effects localised to the lowest 20 km of atmosphere \cite{andrews_laser_2023}. In comparison, downlink channels are substantially less lossy, since by the time the optical signal has reached the turbulent layer the wavefront has expanded in size via diffraction to be greater than the length scales associated with turbulent effects \cite{coulman_outer_1988}. On the other hand, previous results have also shown that the relay-configured protocol, when used in association with quantum amplification \cite{winnel_overcoming_2021}, provides a rate of distribution of entanglement on the order of $\sim\bigO(\eta)$, where $\eta^2$ is the total combined loss of both uplink channels. In comparison, the scaling of the distribution configuration is known to be $\sim\bigO(\eta^2)$. The point we wish to investigate here is therefore this: given a set of realistic atmospheric channels describing an optical path from ground-to-satellite and vice-versa, does the superior scaling of the relay configuration overcome the increased loss of the uplink channel, relative to the reduced loss but poorer scaling of the downlink channel?

We propose to answer this question by examining the performance of both the relay and distribution protocols for three separate cases. In the first case, we examine the performance of three satellites linked by symmetric channels, and document their performance in the relay and distribution configurations. In the second case, we consider the same setup, but now with Alice and Bob permitted to use a first-order quantum scissor NLA \cite{ralph_nondeterministic_2009, xiang_heralded_2010} to amplify any received signals. In the third and last case, we consider the same quantum protocol as in the second case, but now relegate Alice and Bob to stations located on the ground where their channels are no longer symmetric and deterministic but are susceptible to atmospheric effects, including the splitting between uplink and downlink channels mentioned above. We restrict our analysis to the distribution of continuous-variable (CV) and single-rail discrete-variable (DV) entangled resources.

The structure of the paper is as follows. In Section \ref{sec:protocols}, we formally describe the quantum protocols employed by Alice and Bob corresponding to the relay and distribution configuration for each of the three cases. Section \ref{sec:atmos_channels} briefly explains the method by which we characterise the uplink and downlink atmospheric channels in terms of loss. We then proceed to show our results in Section \ref{sec:results}, and discuss the significance of our findings and our conclusions in Section \ref{sec:discussion}.

\section{Protocols} \label{sec:protocols}
In this section, we describe the two fundamental classes of protocols under scrutiny (the relay and distribution configurations respectively). We first consider each configuration when employed in a triple-satellite network, where Alice and Bob transmit and receive with Charlie via a diffraction-limited optical beam propagating through vacuum, which we model as a deterministic pure-loss channel of fixed transmissivity $\eta$. We then consider each configuration over the same network, but now where Alice and Bob have access to first-order quantum scissors, with which they can amplify a quantum state non-deterministically. In the relay case, Bob chooses to perform a distributed NLA \cite{mauron_comparison_2022, winnel_overcoming_2021}, since this is a better choice than na{\"i}vely amplifying the output for both DV and CV. Lastly, we consider a more sophisticated scenario, where Alice and Bob have access to nonlinear amplification but are now instead separated via asymmetric stochastic atmospheric channels of transmissivity $\eta_A$ and $\eta_B$ respectively, corresponding to fading uplink and downlink channels.
\begin{figure*}[!t]
    \centering
    \includegraphics[width = \textwidth]{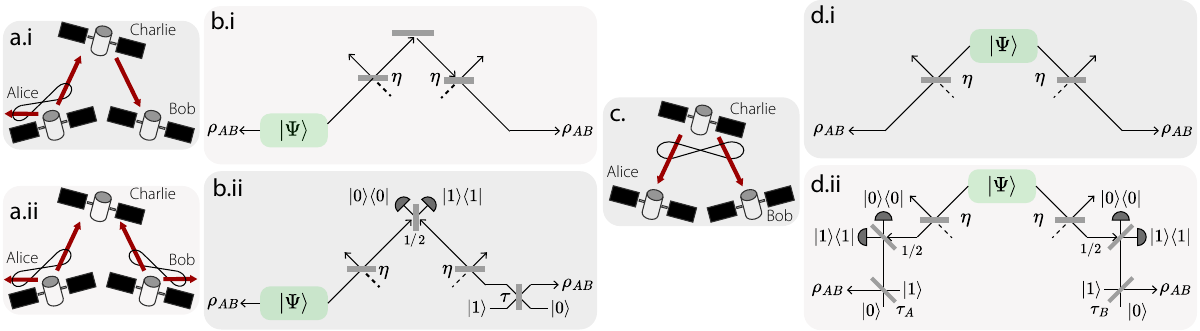}
    \caption{Protocol diagrams for the triple-satellite network, where Alice, Bob and Charlie each operate a satellite in low-Earth orbit. Alice and Bob are connected via the intermediary central satellite Charlie. The optical link between both Alice and Charlie as well as Bob and Charlie is a simple diffraction-limited beam propagating through vacuum, and is assumed to be deterministic, symmetric between Alice and Bob and directionally isotropic. Each link is modelled as a pure-loss bosonic channel of transmissivity $\eta$. In the unamplified relay configuration (\textbf{a.i, b.i.}) Alice generates an entangled resource, one half of which is transmitted to Charlie, who forwards it to Bob unchanged. In the amplified relay configuration (\textbf{a.ii, b.ii.}), Alice and Bob each generate an entangled resource and send half to Charlie, who transfers Alice's entanglement to Bob via entanglement swapping. In the distribution configuration (\textbf{c.}), the entangled resource is generated centrally by Charlie and transmitted to Alice and Bob, who then either do nothing (\textbf{d.i.}) or amplify the received state (\textbf{d.ii.}). \textcolor{black}{Each protocol is technically resource-agnostic ($\ket{\Psi}$ can be continuous- or discrete-variable); we denote the CV (DV) version of a protocol as one that begins with a CV (DV) resource and aims to distill CV (DV) entanglement.}}
    \label{fig:protocol_diagrams_satel}
\end{figure*}

\subsection{Unamplified protocols}
\label{sec:protocols_unamplified}
It is instructive to first consider a simplified scenario wherein Alice and Bob are aboard two satellites, and they wish to distribute entanglement between themselves via two free-space optical links mediated by an intermediary satellite Charlie. The connection between Alice and Charlie is modelled as a simple pure-loss channel $\E_A(\eta_A)$ for fixed transmissivity $\eta_A$. We assume symmetry between the two parties such that the connection between Charlie and Bob is modelled by the same channel, $\E_B(\eta_B) = \E_A(\eta_A)$, and we assume symmetry in the transmissivities of each channel, $\eta_A = \eta_B = \eta$.

The first strategy Alice and Bob may use to distribute entanglement is by generating the entangled resource locally and transmitting it to the opposing party via Charlie, who acts as a relay station (Fig. \ref{fig:protocol_diagrams_satel}.a.i). Assume that Alice generates a two-mode entangled resource state $\ket{\Psi}$ which she intends on sharing with Bob. Alice distributes one arm of the resource $\ket{\Psi}$ by transmitting it to Charlie via $\E_A$, who directly routes it to Bob via $\E_B$ without making any alterations to the signal, for example by reflecting the beam over the horizon via a mirror (Fig. \ref{fig:protocol_diagrams_satel}.b.i).

The second strategy Alice and Bob may use is to generate the entangled resource externally at Charlie, who then distributes each half of the entangled resource to Alice and Bob (Fig. \ref{fig:protocol_diagrams_satel}.c). Assuming now that Charlie generates the two-mode entangled resource state $\ket{\Psi}$, he transmits one arm to Alice via the $\E_A$ while simultaneously transmitting the other arm to Bob via $\E_B$ (Fig. \ref{fig:protocol_diagrams_satel}.d.i).

\subsection{Amplified protocols}
\label{sec:protocols_amplified}
Noiseless linear amplifiers (NLAs) can be used to improve the average performance of both CV and DV entanglement distribution schemes \cite{winnel_generalized_2020, winnel_overcoming_2021, blandino_improving_2012, ghalaii_long-distance_2020, notarnicola_long-distance_2023}. NLAs can noiselessly amplify a quantum signal, but do so non-deterministically so as to avoid violating the no-cloning theorem \cite{wootters_single_1982,dieks_communication_1982,barnum_noncommuting_1996}. We thus consider the two strategies used by Alice and Bob in Section \ref{sec:protocols_unamplified}, except now Alice and Bob may use one or more first-order quantum scissors to amplify the entanglement of their shared state after transmission. A quantum scissor of the first order is a type of NLA that enacts the map $\hat T_1$ \cite{winnel_generalized_2020}
\begin{align}
    \hat T_1 \ : \ket{\phi} \longrightarrow \sqrt{\frac{1 - \tau}{2}} c_0 \ket{0} \textcolor{black}{\pm} \sqrt{\frac{\tau}{2}} c_1 \ket{1}
\end{align}
for a state $\ket{\phi}$ with Fock-basis representation $\ket{\phi}~=~\sum_{n = 0}^{\infty} c_n \ket{n}$ and transmissivity parameter $\tau$. The quantum scissor operates by mixing the target state with the unbalanced Bell state $\ket{\psi} = \sqrt{\tau}\ket{0,1}+\sqrt{1-\tau}\ket{1,0}$ on a balanced beamsplitter and then measuring the two output ports in the Fock basis. Successful operation is heralded by a measurement result showing a single photon in either output port and vacuum in the other \footnote{The specific detector that detects a click is not important; the two successful outcomes only differ by a correctable phase difference of $\pi/2$.}, represented by the POVM outcome 
\begin{align}
\Pi_\text{succ} = \ket{1,0}\bra{1,0}\otimes\mathbb{I} + \ket{0,1}\bra{0,1}\otimes \id
\end{align}
It can be seen that the quantum scissor is essentially a teleportation operation from the general input state $\ket{\phi}$ onto the finite-dimensional resource state $\ket{\psi}$, where $\ket{\phi}$ is also amplified. This is also known as teleamplification \cite{guanzon_ideal_2022}; by defining a gain parameter $g~\equiv~\sqrt{\tau/(1-\tau)} \geq 1$ the output state becomes (up to normalisation)
\begin{align}
    \hat T_1 \ket{\phi} &= \sqrt{\frac{1 - \tau}{2}} c_0 \ket{0} \textcolor{black}{\pm} \sqrt{\frac{\tau}{2}} c_1 \ket{1} \\
    &= \sqrt{\frac{1-\tau}{2}} \left[ c_0\ket{0} \textcolor{black}{\pm} \sqrt{\frac{\tau}{1-\tau}} c_1 \ket{1} \right] \\
    &\simeq c_0\ket{0} \textcolor{black}{\pm} g c_1 \ket{1}
\end{align}
such that for small-amplitude ($c_{2,3, ...} \ll 1$) states the scissor approximates the ideal noiseless amplification map $\ket{\phi} \longrightarrow \sum_{n = 0}^{\infty} c_n g^n \ket{n}$. The probability of a successful amplification is given by
\begin{align}
    p & = \frac{1}{2(1+g^2)} \left[ c_0^2 + g^2 c_1^2 \right].
\end{align}

For the satellite network in the relay configuration, the strategy Alice and Bob use is the protocol introduced in \cite{winnel_overcoming_2021} \textcolor{black}{for CV entanglement, and later extended to DV entanglement in \cite{mauron_comparison_2022}} (Fig. \ref{fig:protocol_diagrams_satel}.a.ii,b.ii). \textcolor{black}{This protocol is a specific case of the more general family of teleamplification protocols \cite{guanzon_ideal_2022}; these are protocols which transmit quantum information indirectly by teleporting an amplified state onto a receiver's mode. For the amplified relay scheme, we use a teleamplification scheme based on a spatially separated first-order quantum scissor of the type described above; the entangled single-rail resource $\ket{\psi}$ is generated by Bob's station and transmitted to Charlie via $\E_B(\eta)$, who performs the non-deterministic photon counting measurement $\hat \Pi_\mathrm{succ}$. Successful measurement heralds a simultaneous teleportation and amplification of Alice's entangled mode $\ket{\Psi}$, sent via $\E_A(\eta)$ to Charlie, onto Bob's retained mode. This protocol, also known as a distributed NLA protocol, has been shown to be superior to direct transmission and localised amplification \cite{mauron_comparison_2022}.}

The distribution configuration for a satellite network employing noiseless amplification is somewhat more straightforward (Fig. \ref{fig:protocol_diagrams_satel}.c, d.ii). The entangled resource $\ket{\Psi}$ is generated by Charlie, who again transmits each half to Alice and Bob via $\E_A(\eta)$ and $\E_B(\eta)$ respectively. Each mode is then input into a local quantum scissor; success is heralded when both Alice and Bob obtain $\Pi_\text{succ}$.

\subsection{Amplified protocols over atmospheric channels}
\label{sec:protocols_atmos}
\begin{figure*}[!t]
    \centering
    \includegraphics[width = \textwidth]{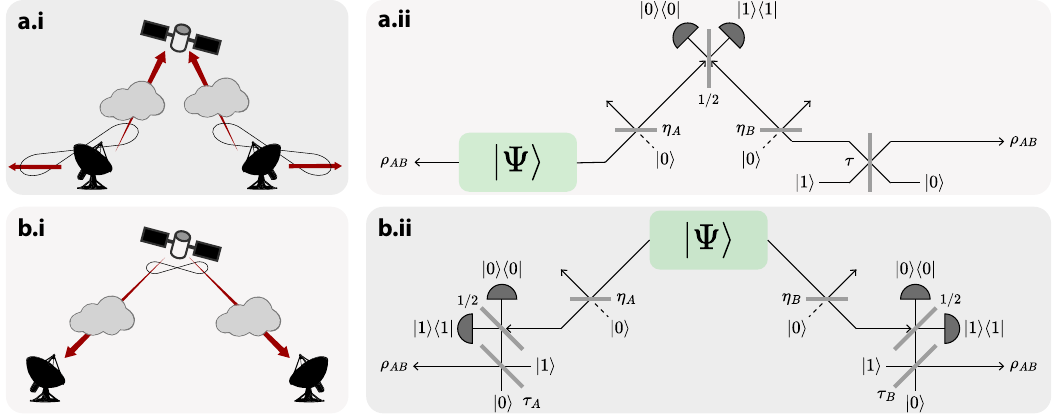}
    \caption{Protocol diagrams for the realistic case of stochastic, asymmetric atmospheric channels. The relay (\textbf{a.i, ii}) and distribution (\textbf{b.i, ii}) protocols are identical to the amplified protocols shown in Fig. \ref{fig:protocol_diagrams_satel}.b.ii and \ref{fig:protocol_diagrams_satel}.d.ii respectively, except the channels linking each station to the central satellite has been replaced with an equivalent pure-loss channel, the transmissivity of which is drawn from a distribution dependent on the atmospheric characteristics of the beam propagation path.}
    \label{fig:protocol_diagrams}
\end{figure*}

One of the key use cases for satellite-mediated entanglement distribution is one where the central satellite is used to mediate the exchange of entangled resources between two ground stations which are spatially separated by a large distance \cite{liao_satellite--ground_2017, ren_ground--satellite_2017}. We therefore consider now a network consisting of two ground stations, Alice and Bob, each communicating with the intermediary low-Earth orbit (LEO) satellite operated by Charlie via an atmospheric free-space optical channel (Fig. \ref{fig:protocol_diagrams}.a.i, b.i). While the overall structure of the relay and distribution protocols remains the same (Fig. \ref{fig:protocol_diagrams}.a.ii, b.ii) as in the triple-satellite network, the channels $\E_A$, $\E_B$ now become asymmetric, anisotropic fading channels with transmission profiles governed by the geometrical and physical characteristics of the propagation path through the atmosphere. We model this by defining the stochastic channels
\begin{align}
    \E_{A,B}^\text{atm} &\equiv \{\E(\eta_{A,B} = \eta)\}, \ \text{where} \\
    \text{prob}(\eta_{A,B} = \eta) &= T^\text{atm}(\eta).
\end{align}
Here $T^\text{atm}(\eta)$ is the probability density function describing the likelihood that the transmission of a state from e.g. Alice to Charlie will experience transmissivity $\eta$. 

The properties of $T^\text{atm}(\eta)$ are dependent on a variety of conditions. Most important is whether the optical path propagates downwards or upwards through the atmosphere, corresponding to a downlink or uplink channel respectively. Other factors include the size of the initial Gaussian profile of the beam; the size of the receiving aperture; the length scales and structure properties of turbulent effects in the atmosphere; environmental factors such as weather and ambient light and the distance and angle of the propagation path. Appendix \ref{sec:atmos_channels} details the procedure used to compute $T^\text{atm}$ and the assumptions made while performing the simulations; we obtain simulated data describing the loss profiles of the uplink and downlink atmospheric channels for a LEO satellite orbiting at 500 km for zenith angles of $\theta = 0^\circ, 10^\circ, 20^\circ$ and $30^\circ$.

\section{Results} \label{sec:results} 
In this section we compute the efficiency of each protocol in generating useful entanglement between Alice and Bob. We quantify the generated entanglement via the rate
\begin{align}
    \R = p\max\{I_{AB}^{\rightarrow}, \ I_{AB}^{\leftarrow}\}
\end{align}
where $p$ is the probability of success of the protocol and $I_{AB}^{\rightarrow}$ ($I_{AB}^{\leftarrow}$) is the (reverse) coherent information\footnote{This expression is unconventional, since in most cases the forward information is trivially less than the reverse and thus usually ignored. We include it here as a generalisation to account for asymmetries that arise from the stochastic channels and amplification steps, since it is assumed that Alice and Bob are able to freely change their classical information flow direction to optimise their distillation in practice.} of the state $\rho_{AB}$ shared by Alice and Bob:
\begin{align}
    I_{AB}^{\rightarrow} = S(\hat \rho_{B}) - S(\hat \rho_{AB}) \\
    I_{AB}^{\leftarrow} = S(\hat \rho_{A}) - S(\hat \rho_{AB})
\end{align}
where $S(\hat \rho)$ is the von Neumann entropy of the state $\hat \rho$. The interpretation of the quantity $\R$ is that of a lower bound on the distillable entanglement \cite{garcia-patron_reverse_2009, weedbrook_gaussian_2012, winnel_generalized_2020} per unit probability. In simpler terms, for Alice and Bob to obtain one e-bit of entanglement, they must exchange on average at most $1/\R$ shots.

As before, we begin by characterising each configuration firstly in terms of the symmetric, unamplified satellite network, secondly in a symmetric satellite network with access to amplification, and lastly in the asymmetric, stochastic ground-satellite-ground network with access to amplification. We show results for the cases in which Alice and Bob seek to distribute both continuous-variable and discrete-variable entanglement.

\subsection{Unamplified protocols}
In the case where Alice wishes to distribute one mode of a single-rail discrete-variable entangled state to Bob via Charlie, she uses the state
\begin{align}
    \ket{\Psi^{DV}} &= \sqrt{\xi} \ket{0,1} + \sqrt{1-\xi} \ket{1,0}
\end{align}
where $\xi$ is a free parameter chosen by Alice. Modelling each channel (Alice to Charlie, Charlie to Bob) as a beamsplitter of transmissivity $\eta$ mixing $\ket{\Psi^{DV}}$ with the vacuum ancilla modes $E_A$, $E_B$, the joint state held by Alice and Bob after transmission $\rho_{AB}$ is
\begin{align}
    \hat \rho_{AB} &= \hat U_{AE_B}(\eta) \hat U_{AE_A}(\eta) \cdot \hat \rho \cdot \hat U_{AE_A}(\eta)^\dagger \hat U_{AE_B}(\eta)^\dagger
\end{align}
for $\hat \rho \equiv \ket{\Psi^{DV}}\bra{\Psi^{DV}}$ and $\hat U_\T{XY}(\eta)$ the unitary operation describing mixing between modes $X,Y$ on a beamsplitter of transmissivity $\eta$:
\begin{align} \label{eq:bs_unitary}
    \hat U_\text{XY}(\eta) \equiv \begin{pmatrix}
        \hat x \\ \hat y
    \end{pmatrix} \longrightarrow \begin{pmatrix}
        \sqrt{\textcolor{black}{\eta}} & -\sqrt{1-\textcolor{black}{\eta}} \\
        \sqrt{1-\textcolor{black}{\eta}} & \sqrt{\textcolor{black}{\eta}}
    \end{pmatrix} \begin{pmatrix}
        \hat x \\ \hat y
    \end{pmatrix}.
\end{align}
Calculating $\hat \rho_{AB}$ is simple:
\begin{align}
    \hat \rho_{AB} &= \begin{pmatrix}
        \xi(1 -\eta^2)  & 0 & 0 & 0 \\
        0 & \eta^2 \xi  & \eta\sqrt{\xi(1-\xi)} & 0 \\
        0 & \eta\sqrt{\xi(1-\xi)} & 1-\xi  & 0 \\
        0 & 0 & 0 & 0 \\
    \end{pmatrix}
\end{align}
It is straightforward to see that in the relay configuration, the reverse coherent information $I_{AB}^\leftarrow$ always exceeds the forward coherent information \cite{garcia-patron_reverse_2009}; we find explicitly then that
\begin{align}
    \R &= p \max\{I_{AB}^\rightarrow, I_{AB}^\leftarrow\} = \max_{\xi \in [0,1]} I_{AB}^\leftarrow \\
    &= \max_{\xi \in [0,1]} 2\xi \tanh^{-1}(1-2\xi)\eta^2 - \mathcal{O}\left(\eta^4\right)\dots
\end{align}
The rate of distribution of entanglement from Alice to Bob in the unamplified symmetric relay configuration is therefore limited by the total channel loss $\eta^2$. Given freedom in the choice of $\xi$, the rate $\R$ is optimised in the high-loss regime by choosing $\xi = 0.217...$, i.e. the value that maximises the quantity $2\xi \tanh^{-1}(1-2\xi)$, and so (Fig. \ref{fig:pubfig_satel}.a)
\begin{align}
    \R &\sim 0.278 \ \eta^2.
\end{align}
Similarly, if instead Charlie distributes $\ket{\Psi^{DV}}$ centrally to Alice and Bob over the same channels, the joint state after transmission is
\begin{align}
    \hat \rho_{AB} &= \hat U_{AE_A}(\eta) \hat U_{BE_B}(\eta) \cdot \hat \rho \cdot \hat U_{BE_B}(\eta)^\dagger \hat U_{AE_A}(\eta)^\dagger \\
    &= \begin{pmatrix}
        1 - \eta & 0 & 0 & 0 \\
        0 & \eta \xi  & \eta \sqrt{\xi (1-\xi)} & 0 \\
        0 & \eta\sqrt{\xi(1 - \xi)} & \eta (1-\xi)  & 0 \\
        0 & 0 & 0 & 0 \\
    \end{pmatrix}.
\end{align}
$\hat \rho_{AB}$ can also be expressed as a mixture of $\hat \rho$ and vacuum:
\begin{align}
    \hat \rho_{AB} &= (1 - \eta)\ket{0}\bra{0} + \eta \ket{\Psi^{DV}}\bra{\Psi^{DV}}.
\end{align}
The distillable entanglement of such a state must necessarily be non-positive when $\eta \leq 1/2$:
\begin{align}
    S(\hat \rho_{AB}) = &-(1-\eta)\log(1-\eta) - \eta\log\eta \\
    S(\hat \rho_{A}) = &-\left[(1-\eta) + \eta \xi\right]\log\left[(1-\eta) + \eta \xi\right] \notag\\
    &- (\eta - \eta\xi)\log\left[\eta - \eta\xi)\right].
\end{align}
When $\eta > 1/2$, any choice of $\xi \geq 0$ drives the eigenvalues of $\hat \rho_A$ closer to the maximal values $\{1/2,1/2\}$, causing an increase in $S(\hat \rho_A)$ relative to $S(\hat \rho_{AB})$ and $I_{AB}^\leftarrow > 0$; in contrast, for $\eta \leq 1/2$ any value of $\xi > 0$ drives the eigenvalues of $\hat \rho_A$ away from the maximal values and so $I_{AB}^\leftarrow \leq 0$. (Note that the symmetry of $\hat \rho_{AB}$ implies $I_{AB}^\leftarrow = I_{AB}^\rightarrow$.) The optimal behaviour of $\xi$ is therefore to be as large as possible ($\xi \rightarrow 1$) for $\eta > 1/2$ and rapidly transition to as small as possible ($\xi \rightarrow 0$) for $\eta \leq 1/2$. However, it is clear to see that for $\eta \leq 1/2$ we find $S(\hat \rho_A) \leq S(\hat \rho_{AB})$ for any $\xi \geq 0$, with equality when $\xi = 0$. This is seen in Figure \ref{fig:pubfig_satel}.a.

In the case where Alice instead wishes to distribute one mode of a continuous-variable two-mode squeezed vacuum (TMSV), she uses the resource \cite{weedbrook_gaussian_2012}
\begin{align}
    \ket{\Psi^{CV}} &= \sqrt{1 - \chi^2} \sum_{n = 0}^\infty \chi^n\ket{n, n}
\end{align}
where the parameter $\chi = \tanh(r)$ is a function of the two-mode squeezing $r \in [0, \infty)$. Because $\ket{\Psi^{CV}}$ does not have a closed expression in the Fock basis, it is not effective to use a density matrix-based approach as in the case of DV entanglement; however, because $\ket{\Psi^{CV}}$ is a Gaussian state transmitted over Gaussian channels $\E_A(\eta)$, $\E_B(\eta)$, the dynamics of the protocol can be compactly represented in the symplectic covariance-matrix formalism without loss of accuracy \cite{weedbrook_gaussian_2012}. In the relay configuration, the joint state after transmission is given by
\begin{align} \label{eq:cv_noamp_relay}
    V_{AB} &\equiv \Tr_{E_A E_B} \LL[ S_{AE_B}(\eta) S_{AE_A}(\eta) V S_{AE_A}(\eta)^T S_{AE_B}(\eta)^T \RR].
\end{align}
Here, $V$ is the covariance matrix of the state $\ket{\Psi^{CV}}$ \cite{weedbrook_gaussian_2012}
\begin{align}
    V &= \begin{pmatrix}
        \nu \id & \sqrt{\nu_2 - 1} \sz \\
        \sqrt{\nu_2 - 1} \sz & \nu \id
    \end{pmatrix}
\end{align}
for quadrature variance $\nu = \frac{1 + \chi^2}{1 - \chi^2} = \cosh(2r)$, identity operator $\id$ and Pauli $Z$ operator $\sz$, and $S_{XY}(\eta)$ is the symplectic map describing the unitary operator \eqref{eq:bs_unitary} \cite{weedbrook_gaussian_2012}. Evaluating \eqref{eq:cv_noamp_relay} yields
\begin{align}
    V_{AB} &= \begin{pmatrix}
        \nu \id & \eta \sqrt{\nu_2 - 1}\sz \\
        \eta \sqrt{\nu_2 - 1}\sz & \left[ 1 + \eta^2(\nu - 1) \right] \id
    \end{pmatrix}
\end{align}
Because the state remains Gaussian, the covariance matrix $V_{AB}$ is sufficient to characterise the entire state and the rate $\R$ can be derived directly from the symplectic eigenvalues of the covariance matrix:
\begin{align}
    I_{AB}^\leftarrow = &\frac{\nu+1}{2}\log\left(\frac{\nu+1}{2}\right) - \frac{\nu-1}{2}\log\left(\frac{\nu-1}{2}\right) \notag\\
    &- \frac{\nu+\eta^2(1-\nu)+1}{2}\log\left(\frac{\nu+\eta^2(1-\nu)+1}{2}\right) \notag\\
    &+ \frac{\nu+\eta^2(1-\nu)-1}{2}\log\left(\frac{\nu+\eta^2(1-\nu)-1}{2}\right) \\
    \R = &\max_{\nu\in[1,\infty)} \frac{\nu+1}{2}\log\left( \frac{\nu+1}{\nu-1} \right)\eta^2 + \mathcal{O}(\eta^4)\dots
\end{align}
The coefficient of the $\eta^2$ term above is monotonic in $\nu$ for the allowed optimisation values, limiting to $1$ for $\nu \rightarrow \infty$:
\begin{align}
    \R \leq \eta^2.
\end{align}
In practice, however, TMSV states of variance $\nu \rightarrow \infty$ require infinite energy, and so we present a practical limit on the rate in the high-loss regime by assuming two-mode squeezing of e.g. $8$ dB (Fig. \ref{fig:pubfig_satel}.b)
\begin{align}
    \R \sim 0.714 \ \eta^2.
\end{align}

Lastly, we consider the case where a two-mode resource is distributed by Charlie along $\E_A(\eta)$, $\E_B(\eta)$ to Alice and Bob. The covariance after transmission is given by
\begin{align}
    V_{AB} &\equiv \Tr_{E_A E_B}\LL[ \hat S_{BE_B}(\eta) \hat S_{AE_A}(\eta) V \hat S_{AE_A}(\eta)^\dagger \hat S_{BE_B}(\eta)^\dagger \RR] \notag \\
    &= \begin{pmatrix}
        \left[ 1 + \eta(\nu - 1) \right] \id & \eta \sqrt{\nu_2 - 1}\sz \\
        \eta \sqrt{\nu_2 - 1}\sz & \left[ 1 + \eta(\nu - 1) \right] \id
    \end{pmatrix},
\end{align}
which is to be expected, given the system is perfectly symmetric. This symmetry implies the forward entanglement distribution capacity is be equal to the reverse entanglement distribution capacity; from \cite{garcia-patron_reverse_2009} and \cite{barnum_quantum_2000} the entanglement distribution capacity assisted by either forward or backward classical communication is equal to the unassisted quantum communication capacity. 

It is known that the single-mode pure-loss channel of transmissivity $\eta$ is antidegradable for $\eta \leq 1/2$ \cite{caruso_degradability_2006}, and so the channels $\mathcal{E}_{A,B}$ each have unassisted quantum communication capacity $\mathcal{Q} = 0$. Thus, defining $\hat \rho' = \mathcal{E}_A(\hat \rho_{CV})$ for $\hat \rho_{CV} = \ket{\Psi_{CV}}\bra{\Psi_{CV}}$, the capacity of the protocol $\mathcal{E}_B(\hat \rho')$ is necessarily zero; hence the capacity of the symmetric protocol $\mathcal{E}_B(\mathcal{E}_A(\hat \rho_{CV}))$ is also zero. Since the RCI lower bounds the distillable entanglement and is $\geq 0$, the RCI is necessarily 0 for $\eta \leq 1/2$.

\subsection{Amplified protocols}

\begin{figure*}[!tb]
    \centering
    \includegraphics[width = 0.49\textwidth]{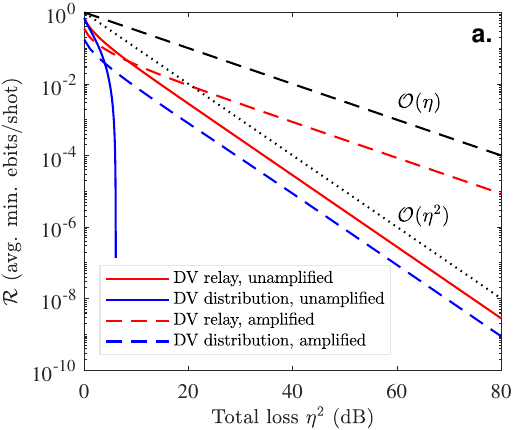}
    \includegraphics[width = 0.49\textwidth]{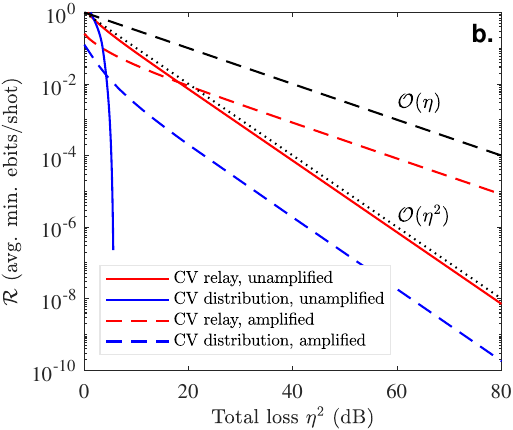}
    \caption{Unamplified (solid lines) and amplified (dashed lines) entanglement distribution protocol rates in the symmetric triple-satellite network. Rates are optimised over all free parameters. \textbf{a.} Distribution of discrete-variable entanglement resource $\ket{\Psi^{DV}}$. \textbf{b.} Distribution of continuous-variable entanglement resource $\ket{\Psi^{CV}}$.}
    \label{fig:pubfig_satel}
\end{figure*}

Calculation of the rate $\R$ is more complicated in the cases where Alice and Bob have access to amplifiers. In the relay configuration, where Alice aims to distribute one arm of an entangled mode to Bob via Charlie, the optimal scheme is not simply to place an amplifier at Bob's station, but rather to perform a \textit{distributed NLA} operation \cite{winnel_overcoming_2021, mauron_comparison_2022}. In the distributed NLA scheme (Fig. \ref{fig:protocol_diagrams_satel}.b.ii), the two stages of the quantum scissor (resource generation and Bell-state measurement) are spatially separated between Bob and Charlie respectively, with Bob transmitting his entangled ancilla mode to Charlie over the channel $\E_B(\eta)$.

For a DV satellite network in this configuration, Alice again begins with the state $\ket{\Psi^{DV}}$, and Bob uses a beamsplitter of variable transmissivity $\tau$ to generate a two-mode entangled ancilla
\begin{align}
    \ket{\phi} &= \sqrt{\tau}\ket{1,0} + \sqrt{1-\tau}\ket{0,1}.
\end{align}
The total joint state prior to transmission is therefore the 6-mode state
\begin{align}
    \ket{\Psi} &= \ket{\Psi^{DV}} \otimes \ket{\phi} \otimes \ket{0,0}
\end{align}
where we label the modes $A_1, A_2$ representing Alice's entangled state $\ket{\Psi^{DV}}$, $B_1, B_2$ representing Bob's entangled ancilla $\ket{\phi}$, and $E_A, E_B$ representing the environment vacuum modes of the channels $\E_A, \E_B$. After Alice and Bob transmit one mode of their respective entangled states to Charlie, he mixes them on a balanced beamsplitter, producing the state
\begin{align}
    \ket{\Psi'} &= 
        \hat U_{A_2B_2}(\tfrac{1}{2}) 
        \hat U_{B_2E_B}(\eta) 
        \hat U_{A_2E_A}(\eta) 
        \ket{\Psi}.
\end{align}
Successful distributed amplification is heralded at Charlie by a single-photon Bell-state measurement outcome of `on' ($\ket{1}\bra{1}$) on one detector and `off' ($\ket{0}\bra{0}$) on the other. This is equivalent to the POVM $\hat \Pi_\mathrm{succ}$ on the modes $A_2, B_2$
\begin{align}
    \hat \Pi_\mathrm{succ} &= \hat \Pi_{A_2 B_2} \\
    &\equiv \ket{0,1}\bra{0,1}_{A_2 B_2} \otimes \id 
    + \ket{1,0}\bra{1,0}_{A_2 B_2} \otimes \id
\end{align}
The joint state after transmission is therefore
\begin{align}
    \hat \rho_{AB} &= \frac{1}{p}\Tr_{A_2B_2E_AE_B}\left[ \hat \Pi_\text{succ} \cdot \hat \rho'\right]
\end{align}
for $\hat \rho' = \ket{\Psi'}\bra{\Psi'}$ and probability of success
\begin{align}
    p &= \Tr[\hat \rho_{AB}].
\end{align}
The forward and reverse coherent information is then calculated directly from the density matrix. Finding the rate $\R$ now requires optimisation over both the initial resource state as well as the gain $g = \sqrt{\tau/(1-\tau)}$:
\begin{align}
    \R &= \max_{\substack{\xi\in[0,1] \\ g\in[1,\infty)}} p \max\{I_{AB}^\rightarrow, I_{AB}^\leftarrow \}
\end{align}

In the distribution configuration for a satellite network, Alice and Bob each have two separate amplifiers to amplify the incoming signal after it is transmitted through the channels $\E_A$ and $\E_B$ respectively (Fig. \ref{fig:protocol_diagrams_satel}.d.ii). Alice and Bob therefore both need separate two-mode ancillae $\ket{\phi_A}$ and $\ket{\phi_B}$, each associated with a separate gain parameter $g_{A,B} = \sqrt{\tau_{A,B}/(1 - \tau_{A,B})}$. For a DV network, the 8-mode initial joint state is
\begin{align}
    \ket{\Psi} &= \ket{\Psi^{DV}} \otimes \ket{\phi_A} \otimes \ket{\phi_B} \otimes \ket{0,0}
\end{align}
where the modes are now labelled $A_1, B_1$, representing Charlie's initial entangled state $\ket{\Psi^{DV}}$, $A_2,A_3$ representing Alice's ancilla $\ket{\phi_A}$; $B_2,B_3$ representing Bob's ancilla $\ket{\phi_B}$, and $E_A. E_B$ representing the environment vacuum modes of the channels $\E_A, \E_B$. Charlie transmits $\ket{\Psi^{DV}}$ to Alice and Bob via $\E_A(\eta), \E_B(\eta)$. Alice (Bob) then amplifies the incoming mode by mixing it on a balanced beamsplitter with the $A_2$ ($B_2$) mode of the ancilla state, producing
\begin{align}
    \ket{\Psi'} &= 
        \hat U_{B_1B_2}(\tfrac{1}{2}) 
        \hat U_{A_1A_2}(\tfrac{1}{2}) 
        \hat U_{B_1E_B}(\eta) 
        \hat U_{A_1E_A}(\eta) 
        \ket{\Psi}
\end{align}
Success is heralded for Alice and Bob individually obtaining successful local amplification, i.e. measuring
\begin{align}
    \hat \Pi_\mathrm{succ} &= \hat \Pi_{A_1A_2} \otimes \hat \Pi_{B_1B_2}
\end{align}
on the output modes of the NLAs. The joint state after transmission is then
\begin{align}
    \hat \rho_{AB} &= \frac{1}{p} \Tr_{A_1B_1A_2B_2E_AE_B} \left[ \hat \Pi_\mathrm{succ} \cdot \hat \rho' \right] \\
    p &= \Tr\left[ \hat \rho_{AB} \right]
\end{align}
and the rate $\R$ is obtained by optimisation of the forward and reverse coherent information over $\xi, g_A$ and $g_B$:
\begin{align}
    \R &= \max_{\substack{\xi\in[0,1] \\ g_A, g_B\in[1,\infty)}} p \max\{I_{AB}^\rightarrow, I_{AB}^\leftarrow \}.
\end{align}
Equivalent results for the CV case are much harder to obtain analytically. Because CV states are by definition infinite-dimensional, and the POVM measurements $\hat \Pi$ are non-Gaussian, exact representations can only be obtained through a Wigner or characteristic function approach (e.g. \cite{zaunders_quantum-amplified_2024, notarnicola_long-distance_2023, ghalaii_long-distance_2020, blandino_improving_2012}) which is extremely computationally intensive. Nevertheless, we can still obtain accurate results by instead considering how the states evolve in a truncated Fock basis. Under this simplification, the amplified CV protocols are approximated by replacing the Alice's infinite-dimensional input two-mode entangled state with an equivalent entangled resource that has been truncated to some finite dimension, i.e.
\begin{align}
    \ket{\Psi^{CV}} &= \sqrt{1-\chi^2} \sum_{n = 0}^N \chi^n \ket{n,n}
\end{align}
The optimised rate $\R$ is then computed directly from the approximate $N$-dimensional output density matrix in exactly the same way as in the DV protocols (substituting $\ket{\Psi^{DV}}$ for $\ket{\Psi^{CV}}$). \textcolor{black}{This is equivalent to working with a truncated series expansion expressed in terms of powers of the squeezing parameter $\chi$, which is accurate so long as the squeezing $\chi$ remains small enough for the higher-order $\sim\mathcal{O}(\chi^{N+1})$ terms to be negligibly small. In this work, the largest value of $\chi$ considered is approximately $\chi_\mathrm{max} \simeq 0.35$; choosing truncation up to $N = 5$ therefore provides at most a relative error of 0.0003\% compared to the full continuous-variable treatment, and so we are confident the truncated model is suitably accurate.} Symbolic calculations were performed using the \textsc{MENTAT} package for Mathematica 14.1 \cite{nicholas_zaunders_nicholaszaundersmentat_2025}. 

In the interests of brevity, we do not present the probabilities of success or density matrices corresponding to the amplified protocols here. The optimised rates $\R$ of the relay and distribution protocols for both DV and CV resources are shown in Figure \ref{fig:pubfig_satel}.a and \ref{fig:pubfig_satel}.b respectively. To summarise, our model of the relay protocol (i.e. the distributed NLA scheme) replicates previous results found for both DV \cite{mauron_comparison_2022} and CV \cite{winnel_overcoming_2021} resource states, where protocol performance is proportional to the probability of success and scales with only the loss $\eta$ of a single channel, not with the entire channel loss $\eta^2$. Further, the amplified distribution scheme scales proportionally with the total loss $\eta^2$ for both DV and CV, though the absolute rate of the CV protocol is lower than the DV protocol. This is a reasonable result: the rate of distribution still scales with the probability of success, as in the relay protocols, but the two amplifiers needed for the distribution protocol mean both detectors must herald success simultaneously and independently to get a successful result. Since the probability of success $p$ for a single channel with a single amplifier scales with the loss of that channel $\eta$, the total probability of success scales as $p^2 \sim \eta^2$ regardless of the resource state. The reduced absolute rate of the CV protocol is ascribed to the truncating effect of the quantum scissor amplifiers, which discards the entanglement stored in the higher degrees of freedom of the input CV state; the DV state experiences no such compression and so generates entanglement at a higher net rate.

\subsection{Amplified protocols over atmospheric channels}

\begin{figure*}[!tb]
    \centering
    \includegraphics[width = 0.475\textwidth]{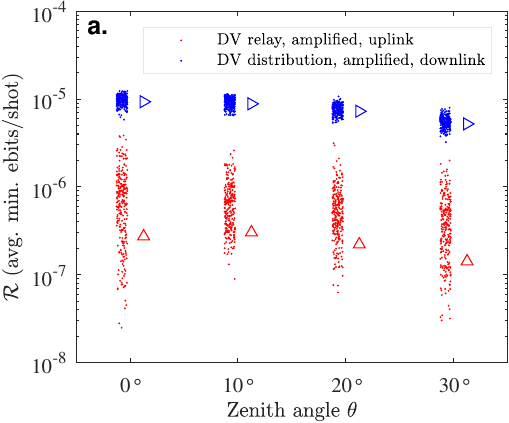}
    \hspace{1em}
    \includegraphics[width = 0.475\textwidth]{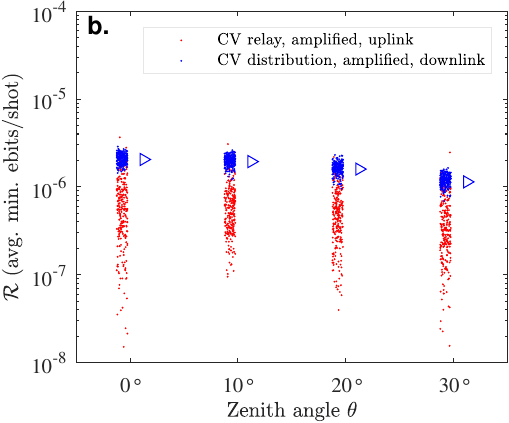}
    \caption{Amplified entanglement distribution protocol rates for the relay and distribution configurations across a ground-satellite-ground network. \textcolor{black}{We show both the rate $\mathcal{R}_\mathrm{ens}$ when considering the mixed state $\hat \rho_\mathrm{ens}$ generated by postselecting based on successful heralded amplification (triangles), as well as the distribution of the reverse information over $n = 250$ random simulations of each protocol over the stochastic atmospheric channels $\E_A(\eta_A)$, $\E_B(\eta_B)$, where $\eta_A,\eta_B$ are drawn randomly from the uplink or downlink channel probability density functions in Figure \ref{fig:pubfig_atmos_multiplot}.c.} \textbf{a.} Distribution of discrete-variable entanglement resource $\ket{\Psi^{DV}}$. \textbf{b.} Distribution of continuous-variable entanglement resource $\ket{\Psi^{CV}}$. \textcolor{black}{Notably, the CV relay protocol is unable to deliver a positive ensemble rate, despite showing positive distillable entanglement on a shot-by-shot basis.}}
    \label{fig:pubfig_atmos_dist}
\end{figure*}

Lastly, we consider the scenario in which Alice and Bob are now confined to ground stations and can only communicate with Charlie, who is located in a satellite at an orbital height of 500 km. In this scenario, Alice and Bob use the same amplified protocols described in the previous section, except now the channels linking Alice and Bob to Charlie take on the fading properties of the atmospheric channels calculated in Section \ref{sec:atmos_channels}.

To simulate each protocol, a random value of $\eta_A$ and $\eta_B$ are drawn from the appropriate distribution $T^\mathrm{atm}(\eta)$ encoding the twin uplink or downlink atmospheric channels $\E_A^\mathrm{atm}, \E_B^\mathrm{atm}$ for either the relay or distribution configurations. The reverse and forward coherent information is then calculated assuming the instantaneous channels $\E_A^\mathrm{atm}(\eta_A), \E_B^\mathrm{atm}(\eta_B)$ and maximised over all free parameters. This is repeated 100 times to provide an estimate of the distribution of $\R$ for the sharing of DV and CV resources in the two competing configurations.

\textcolor{black}{We also consider the entanglement in the bulk (asymptotic) case, since it is a known result that channel fluctuations arising from atmospheric effects give rise to an effective noise that reduces the purity of the shared entangled state \cite{hosseinidehaj_composable_2021}. Assuming that Alice and Bob mutually discard states which coincide with measurement heralding a failed result, the resulting ensemble state is the mixture
\begin{align}
    \hat \rho_\mathrm{ens} &= \int \dd \eta_A \ \dd\eta_B \ T^\mathrm{atm}(\eta_A, \eta_B) \ p(\eta_A, \eta_B) \ \hat \rho_{AB}
\end{align}
and so we define the ensemble rate $\mathcal{R}_\mathrm{ens}$
\begin{align}
    \mathcal{R}_\mathrm{ens} &= \max_{\xi, \chi \dots} \max \{ I_{AB}^\leftarrow, I_{AB}^\rightarrow \}
\end{align}
where the optimisation $\max_{\xi,\chi,\dots}$ is performed over the same free parameters for each resource type and configuration as in the non-ensemble case, i.e. $\xi \in [0,1]$ and $\chi \in [0,\chi_\mathrm{max})$ for DV and CV respectively and $g \in [1,\infty)$ and $g_A,g_B \in [1,\infty)$ for the relay and distribution configuration respectively.
}

Figure \ref{fig:pubfig_atmos_dist} shows these results for zenith angles $\theta$ of $0^\circ$ (directly overhead), $10^\circ$, $20^\circ$ and $30^\circ$, mimicking the arc of a satellite as it passes above a station. The most immediate observation is that the distribution of the potential rates closely reflects the distribution of the input channel transmissivities, with the rate of the distribution protocol exhibiting a low variance commensurate with the diffraction-limited downlink channels. Similarly, the rates achieved by the relay protocol span an interval of several orders of magnitude, as might be expected from the extreme fading seen for atmospheric uplink channels. Despite these differences, the two protocols are comparatively evenly matched in performance for the given atmospheric path and beam characteristics, especially for the CV case. This is not unreasonable to expect, however, given the loss profiles seen in Figure \ref{fig:pubfig_atmos_multiplot}.c and the expected loss scaling of the relay and distribution configurations: a crude order-of-magnitude analysis shows at $Z = 500$ km and $\theta = 0^\circ$ the amplified distribution protocol generates entanglement at a rate proportional to $\eta^2$ over channels of loss $\eta \sim 20$ dB, whereas the relay protocol generates entanglement at a rate proportional to $\eta$ but over channels which achieve $\eta \sim 40$ dB (albeit with low probability). Hence both protocols obtain roughly $\R \sim 10^{-4}$. Despite this, the high variability of the uplink protocol means that for low-Earth orbits the average rate $\R$ would be much lower than the equivalent downlink configuration, and so it is clear that in practice the distribution protocol is the most appropriate choice.

Furthermore, Figure \ref{fig:pubfig_atmos_multiplot} also highlights a trend in performance between the discrete-variable (Fig. \ref{fig:pubfig_atmos_multiplot}.a) and continuous-variable (Fig. \ref{fig:pubfig_atmos_multiplot}.b) distribution protocols. While there is comparatively little difference between the choice of DV or CV resources when using the relay protocol, the distribution protocol shows up to an order of magnitude improvement when distributing DV entangled resources as compared to CV. We ascribe this difference to the truncating effect of the quantum scissors. Continuous-variable entangled states of high squeezing are generally capable of holding more `raw entanglement' compared to maximal-entanglement DV states, as evidenced by the difference in $\R$ between the unamplified DV distribution scheme and the unamplified CV distribution scheme (Fig. \ref{fig:pubfig_satel}.a and \ref{fig:pubfig_satel}.b respectively), by exploiting the additional degrees of freedom associated with the continuous-variable Hilbert space; however, this advantage becomes a disadvantage when paired with truncating quantum amplifiers such as the quantum scissor. When amplification is performed on a CV state, the entanglement stored in the higher degrees of freedom is lost on truncation, leading to lower accessible entanglement overall compared to the DV state, which starts off with less entanglement but loses none as a result of amplification. Overall, it can be concluded that a setup in which Alice and Bob share discrete-variable entanglement over a distribution configuration is likely the optimal choice for satellite-mediated entanglement distribution between distant ground stations.

\textcolor{black}{In terms of the ensemble results, we identify that the bulk distillable entanglement $\mathcal{R}_\mathrm{ens}$ behaves as expected given the distribution of the information of the states shot-by-shot and the general non-additivity of quantum resources \cite{garcia-patron_reverse_2009}. That is, the quantum noise penalty associated with the channel fluctuations on an ensemble level is apparent as a decrease in the ensemble distillable entanglement, relative to the mean `centroid' value of the protocol entanglement on a shot-by-shot basis. This effect has also been reported in CV-QKD performed over atmospheric channels, where the fluctuation of the channel generates an effective covariance noise term proportional to the variance of the channel distribution $T^\mathrm{atm}(\eta)$ \cite{hosseinidehaj_composable_2021}. This effect is unsurprisingly most noticeable in the relay protocols, where the high-variability uplink channels greatly degrade the ensemble distillable entanglement, and negligible in the distribution protocols, where the localised channel distribution leads to a negligible noise term. Interestingly, while the DV relay protocol does manage to produce positive ensemble distillable entanglement (i.e. the protocol could feasibly be used to distil entangled resources), the CV relay protocol cannot. This is consistent with CV resources being generally less robust to loss compared to DV \cite{dias_distributing_2022}, at least in the regime of idealised operation and state generation.}

Several other practical factors recommend the distribution protocol as well: for example, the channel statistics of a downlink channel can be easily improved by increasing aperture size, which cannot be readily done on a satellite, and downlink channels can crucially use adaptive optics techniques to greatly decrease atmospheric distortion. Similarly, quantum amplification (which is currently still limited to laboratory implementation) would also be much easier to conduct within the confines of a ground station.

\section{Discussion} \label{sec:discussion}

In this work we have studied the feasibility of several methods of entanglement distribution via satellite. We considered two use cases relevant to future quantum networks: the first being distribution of an entangled resource between two distant satellites via a central satellite, the second being distribution between two ground stations spatially separated over a large distance via a satellite in low-Earth orbit. We then consider two competing protocols which can distribute entangled states over such networks; firstly, a protocol which uses the intermediary satellite as a relay, and secondly a protocol which distributes entanglement directly from the satellite to each station. 

We show that, for the satellite-satellite-satellite network which assumes vacuum propagation and symmetric, deterministic, pure-loss channels, the relay configuration is undoubtedly superior to the distribution configuration. We quantify the performance of each protocol explicitly in terms of the forward and reverse coherent information, which lower-bounds the distillable entanglement of the protocol. Our analysis shows the optimised relay protocol distributes entanglement at a rate that scales quadratically with individual channel loss, whereas the optimised distribution scheme is unable to distribute entanglement above 3 dB of single-channel loss. We demonstrate this outcome for the distribution of both discrete-variable (Bell-state) and continuous-variable (two-mode squeezed vacuum state) resources.

Secondly, we consider the effect of noiseless linear amplification on each configuration under the satellite network described above. For the relay configuration, we model the distributed NLA protocol proposed by \cite{winnel_overcoming_2021}; for the distribution configuration, we assume separate receiver-side amplification at both ground stations. Under these conditions, the relay protocol distributes entanglement at a rate linear with channel loss when optimised for input state and amplifier gain. The distribution protocol distributes entanglement at a rate quadratically with loss. Again, we show these results hold for both CV and DV resources.

Thirdly, we model the performance of the amplified protocols for the ground-satellite-ground network, where we accurately represent the properties of free-space optical channels over ground-to-satellite or satellite-to-ground transmission. We obtain these channels by simulating the propagation of a Gaussian optical beam through an atmospheric channel via a split-step method using phase screens and we generate simulated data characterising the probability density functions of channel transmissivity for both uplink and downlink transmission for a satellite in low-Earth orbit. When we consider the performance of the amplified relay protocol across twin stochastic uplink channels, compared with the amplified distribution protocol across twin stochastic downlink channels, we see that for a satellite at an orbital height of 500 km the two protocols are roughly equal in the rate at which they distribute entanglement. However, the strong fading characteristics of the uplink channel make it overall a poor choice for entanglement distribution, with the relay protocol only exceeding the distribution protocol a small fraction of the time. Lastly, we quantify our findings for both CV and DV resources, and show conclusively that while the higher-loss relay configuration is not highly dependent on the source entanglement, the lower-loss distribution configuration performs markedly better for discrete-variable entanglement distribution. We therefore conclude that out of the presented options, distributing entanglement between two ground stations via satellite is likely best achieved by generating single-rail-entangled Bell pairs centrally and distributing them via atmospheric downlink channels before amplifying with first-order quantum scissors.

\textcolor{black}{Lastly, we suggest a number of ways to improve the accuracy and practical value of the results shown above with respect to satellite communication. A key limitation of the analysis presented here are the technical assumptions made on the atmospheric channels $\E_{A,B}^\mathrm{atm}$, and the practical consideration required when transmitting and receiving quantum states sent over free-space channels. In particular, noise sources arising from non-ideal channels or detection can have a significant effect on the quality of entanglement obtained by Alice and Bob after transmission. With regards to the accuracy of the channel assumptions, requiring night-time operation is sufficient to justify the use of pure-loss channels; however, in the case of day-time operation contributions from thermal effects cannot be ignored. Spectral, spatial or temporal filters centred on the signal wavelength can be used to effectively mitigate the effect of background thermal noise \cite{cao_long-distance_2020, liao_satellite--ground_2017}. It is also worth pointing out that daytime operation induces much stronger turbulent effects by virtue of direct solar heating, and would probably magnify the contrast between downlink and uplink channels even further in terms of variability and mean loss. Additionally, there are a slew of further losses and non-Gaussian noise sources which we do not include here which must nevertheless be accounted for in practical deployment of any satellite-based protocol, which arise from practical limitations of the satellite platform and free-space detection. For example, a challenge of free-space optics is maintaining indistinguishability of spatial modes between transmitted states which must be interfered, such as when Charlie performs a Bell-state measurement for the relay protocol. The usual solution requires spatial filtering via single-mode fibre \cite{cao_long-distance_2020}, which is sufficient but also introduces a further coupling loss. Adaptive optics can also be used to correct spatial distortions to assist with single-mode fibre coupling \cite{cao_long-distance_2020}, in addition to reducing channel losses overall. Additional noise sources include phase noise, which can arise due to the need for phase stability in single -rail systems; in this case using a digital local oscillator which is reconstructed at Charlie and Bob can help mitigate the total noise \cite{pirandola_satellite_2021}. Satellites in the LEO region which move quickly relative to the ground station are also susceptible to significant Doppler shifting, which can affect both clock timing and the central frequency of the transmitted mode. Postprocessing methods can alleviate these errors associated with clock timing on the practical layer \cite{takenaka_satellite--ground_2017}, but addressing the phase noise and frequency drift from Doppler effects is more difficult. We direct the reader to the work of Vallone et al. \cite{vallone_interference_2016} on Doppler noise sources in single-photon interference, and note that Wu et al. \cite{wu_single-photon_2024} provide a detailed treatise on the mitigation of phase errors in single-photon atmospheric transmission via a variety of techniques including unbalanced interferometry and careful correction of measurable Doppler effects via controllable wave plates.}

\vspace{1em}
\begin{acknowledgments}
\vspace{-1em}
The Australian Government supported this research through the Australian Research Council’s Linkage Projects funding scheme (Project No. LP200100601). The views expressed herein are those of the authors and are not necessarily those of the Australian Government or the Australian Research Council. We also acknowledge partial support from Northrop Grumman Corporation and useful discussions with Ryan Aguinaldo. \cite{zaunders_entanglement_2025}
\end{acknowledgments}

\appendix
\section{Atmospheric channel modelling} \label{sec:atmos_channels}

We simulate the properties of the uplink and downlink atmospheric channels via a split-step beam propagation method, combined with a Kolmogorov turbulence model to produce stochastic phase screens that mimic random atmospheric effects, based off the methods detailed in \cite{schmidt_numerical_2010}, \cite{sasiela_electromagnetic_2007} and \cite{villasenor_alvarez_advanced_2023}. The phase screen approach approximates the propagation of an optical field through a turbulent medium, such as the atmosphere, by segmenting the propagation path into $n$ discrete volumes of variable width $\Delta h_i$, such that $\Delta h_1 + \Delta h_2 + ... + \Delta h_n = Z$ for the total propagation distance $Z$. To account for the effects of stochastic turbulence, the optical effect of each segment is encoded into a two-dimension screen of randomly-generated phase values whose power and spectral density encode the turbulent characteristics of the entire volume. At each step in the propagation, the algorithm approximates non-vacuum propagation through the segment by mixing the optical field with the generated phase screen and then propagating the resulting field through a vacuum beam path of equal length to the segment. At the end of the propagation path, the receiver plane encodes the full phase and intensity information of the propagated beam; the transmission $\eta$ of the beam is given by the ratio of the received power to the input power
\begin{align}
    \eta &\equiv \frac{\int_{\abs{\bs{x_n}}^2 \leq a_r^2} \abs{\Phi_\text{rec}}^2 \dee^2 \bs{x_n}}{\int \abs{\Phi_\text{source}}^2 \dee^2 \bs{x_1}}.
\end{align}
Here $\Phi_\text{source}$ is the optical field prior to propagation and $\Phi_\text{rec}$ the optical field received at the aperture described in terms of the initial and received Fourier plane coordinates $\bs{x_1}, \bs{x_n} \in \mathbb{R}^2$ respectively. The radius of the receiver aperture is signified by $a_r$.

The procedure for generating each altitude-dependent phase screen begins with a choice of atmospheric refractive-index structure parameter $C_n^2$. The model used in this work is the conventional Hufnagel-Valley 5/7 model
\begin{align}
    C_n^2(h) = \ &5.94\times10^{-53}\left(\frac{v}{2}\right)^2 h^{10} e^{-h/1000} \notag \\
    &+ 2.70\times10^{-16} e^{-h/1500} + Ae^{-h/100}
\end{align}
with high-altitude wind pseudospeed $v = 27$ ms$^{-1}$ and surface value $A = 1.7\times10^{-14}$ m$^{-2/3}$ \cite{fiorino_propagation_2022, sasiela_electromagnetic_2007, chahine_beam_2020}. The position of each phase screen is determined by choosing $n + 1$ points $h_i$ along the propagation path $Z$, with $h_0 = 0$ and $h_{n+1} = Z$, such that for each atmospheric slice of width $\Delta h_i = h_{i} - h_{i-1}$ the net addition to the cumulative scintillation index (measured by the Rytov parameter $r_R^2$) is not greater than 0.2 \cite{villasenor_alvarez_advanced_2023, martin_intensity_1988}, i.e.
\begin{align}
    r_R^2 = 1.23 k^{7/6} \int_{h_{i-1}}^{h_i} C_n^2(h) \left( h - h_{i-1} \right)^\frac{11}{6} \dee h \leq 0.2. \label{eq:equal-rytov}
\end{align}
Here $k = 2\pi/\lambda$ is the wavenumber of the propagating field, $\lambda$ the wavelength and $\zeta$ the zenith angle of the propagation path relative to the ground station \footnote{The use of $\sec(\zeta)$ is an approximation only valid for $\zeta \leq 1$ rad of angle. At steeper angles, modifications must be made to account for optical path distortion and higher atmospheric effects \cite{gonzalez-raya_satellite-based_2024}. However, this work only considers shallow zenith angles.}.

An important note is that the HV-5/7 model only predicts significant turbulent effects ($C_n^2 \geq 10^{-20}$) below altitudes of approximately 20 km. Because of this, the above method fails to work outside this region, since the amount of atmosphere required to obtain a scintillation variance contribution of 0.2 for $h \geq 20$ km is larger than the propagation distance $Z$ for LEO. The final phase screen volume $\Delta h_n$ therefore accounts for the entire propagation path between $h_n \sim 20$ km and $h_{n+1} = Z$.

The next step is to calculate the Fried coherence parameter $r_0$ for the uplink and downlink directions respectively. In the case of downlink, the relative lack of turbulence in the atmosphere at high altitudes means the primary loss mechanism for a Gaussian beam is simple diffraction \cite{gonzalez-raya_satellite-based_2024, pirandola_satellite_2021, schmidt_numerical_2010}; by the time the beam has reached the upper bound of the turbulent zone at $h = 20$ km the wavefront has expanded to be larger than the length scales at which turbulent effects operate and is well approximated by a plane wave undergoing diffractive loss. The Fried coherence can therefore be written as
\begin{align}
    r_0^\text{downlink} = \left[ 0.423 k^2 \int_{h_{i-1}}^{h_i} C_n^2(h) \dee h \right]^{-3/5}.
\end{align}
In the case of an uplink channel, the beam instead encounters the regions of highest turbulence at a point where it is small and has a high curvature, rendering it highly susceptible to turbulent eddies on both slow and fast time scales, leading to beam wandering and beam broadening effects respectively \cite{gonzalez-raya_satellite-based_2024}. The calculation of the Fried coherence is therefore slightly more involved \cite{villasenor_alvarez_advanced_2023, andrews_laser_2023}:
\begin{align}
    r_0^\text{uplink} = \left[ 0.424 k^2 \left( \mu_1 + 0.622 \mu_2 \Lambda^{11/6} \right) \right]^{-3/5}
\end{align}
where $\mu_1, \mu_2$ are the altitude moments of the structure parameter \cite{sasiela_electromagnetic_2007}
\begin{align}
    \mu_1 &= \int_{h_{i-1}}^{h_i} C_n^2(h) \left( \frac{\Theta (Z - h)}{Z} + \frac{h}{Z} \right)^{5/3} \dee h \\
    \mu_2 &= \int_{h_{i-1}}^{h_i} C_n^2(h) \left( 1 - \frac{h}{Z} \right)^{5/3} \dee h
\end{align}
for $\Theta, \Lambda$ the output-plane Gaussian beam parameters \cite{andrews_laser_2023} respectively
\begin{align}
    \Theta &= 1 + \frac{Z}{R} \\
    \Lambda &= \frac{2 Z}{k \omega}
\end{align}
and
\begin{align}
    R &= Z \left[ 1 + \left( \frac{\pi \omega_0^2}{\lambda Z} \right)^2 \right] \\
    \omega &= \omega_0 \sqrt{1 + \left( \frac{\lambda Z}{ \pi \omega_0^2} \right)^2}.
\end{align}
Here $\omega_0$ is the initial Gaussian beam waist and $R$, $\omega$ the radius of curvature and spot size of the Gaussian beam at distance $Z$ respectively.

The coherence diameters $r_0$ of the beam in each slice govern the power spectral density used to generate the random phase screens used in the propagation algorithm. The modified von K{\'a}rm{\'a}n model of spectral density is used here \cite{villasenor_alvarez_advanced_2023, schmidt_numerical_2010}:
\begin{align}
    \Phi_\phi^\text{mvK}(\kappa) &= 0.49 r_0^{-5/3} \frac{e^{-\kappa^2/\kappa_m^2}}{(\kappa^2 + \kappa_0^2)^{11/6}}.
\end{align}
The angular spatial frequency values $\kappa_m = 5.92/\ell_0$ and $\kappa_0 = 2\pi / L_0$ encode the small-scale, high-frequency and large-scale, low-frequency characteristics of the turbulent medium; these values are derived from a Coulman-Vernin atmospheric model \cite{coulman_outer_1988}
\begin{align}
    L_0 &= \frac{25\times10^6}{6.25\times10^6 + (h - 8500)^2}, \\
    \ell_0 &= 0.005 L_0.
\end{align}
The power spectral density $\Phi_\phi^\text{mvK}$ dictates the distribution of the Fourier series coefficients corresponding to each phase screen $\phi(x_i, y_i)$. An array of random values are drawn from $\Phi_\phi^\text{mvK}$ for each point in the discretized 2D screen and converted to random phase values via Fourier transform. The accuracy of each phase screen is additionally improved via the subharmonic method \cite{lane_simulation_1992}.

\begin{figure*}[!t]
    \centering
    \includegraphics[width=\linewidth]{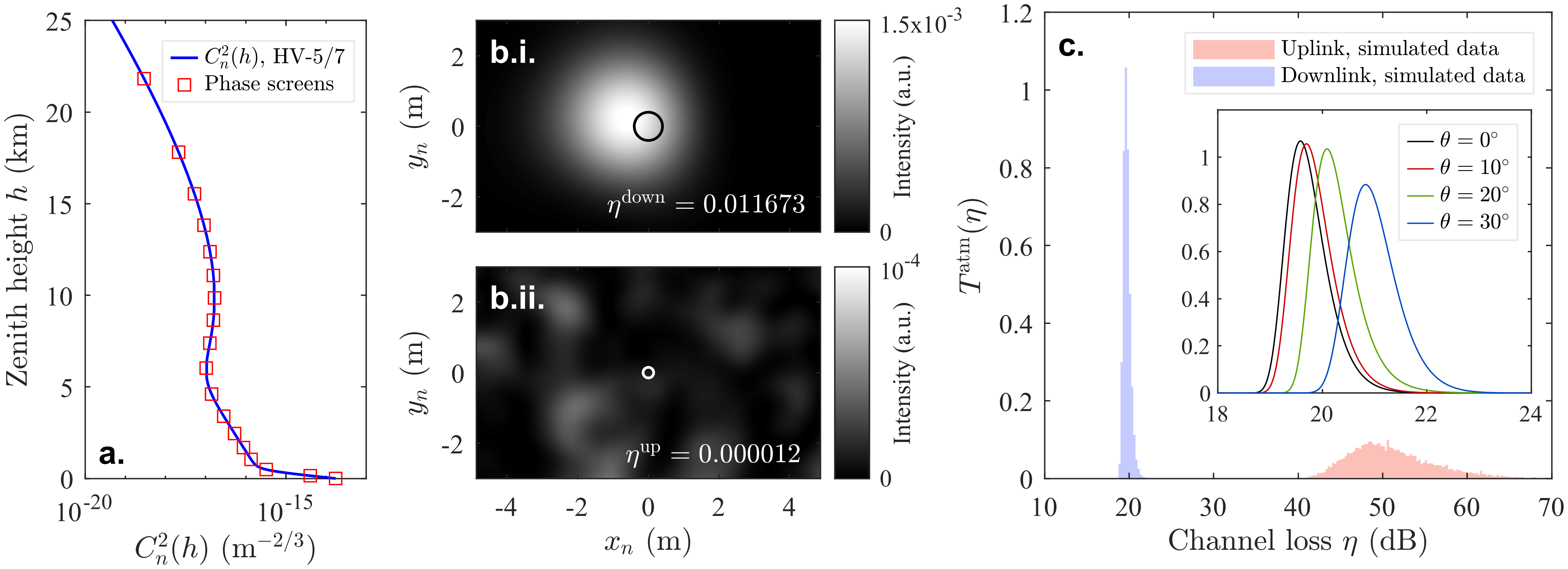}
    \caption{\textbf{a.} The atmospheric structure constant $C_n^2$ for the Hufnagel-Valley 5/7 model as a function of altitude $h$ for zenith angle $\theta = 0$ rad. Propagation of an optical field through the atmospheric medium is emulated via discrete phase screens, which encode the turbulent characteristics of the atmospheric volume between screens. The phase screen locations (boxes) are selected according to the equal-Rytov condition \eqref{eq:equal-rytov}. \textbf{b.} Optical field intensity $\abs{\Phi_\mathrm{rec}}^2$ at the receiver plane with real-space coordinates $(x_n, y_n)$ for downlink (\textbf{i.}) and uplink (\textbf{ii.}) transmission. The initial beam profile is a Gaussian beam with beam waist $\omega_0 = 0.1$ m and wavelength $\lambda = 1550$ nm. Circles denote the receiver apertures of radius $a_r = 0.4$ m (0.15 m) in the receiver plane for downlink (uplink). The transmissivity $\eta^\mathrm{down}$ ($\eta^\mathrm{up}$) is given by the ratio of received power (received intensity integrated over the aperture) to initial power. \textbf{c.} Distribution of $T^\mathrm{atm}(\eta)$ of the downlink and uplink atmospheric channels for zenith height $Z = 500$ km, zenith angle $\theta = 0$ rad over $N = 10 000$ simulated propagations. Inset: fitted $T^\mathrm{atm}(\eta)$ for downlink channels with zenith height $Z = 500$ km and zenith angle $\theta = 0^\circ, 10^\circ, 20^\circ$ and $30^\circ$.}
    \label{fig:pubfig_atmos_multiplot}
\end{figure*}

In the simulations presented in this work, we consider a coherent Gaussian beam with frequency $\lambda = 1550$ nm and initial beam waist $\omega_0 = 10$ cm. The beam is transmitted between a ground stattion at sea level $h = 0$ m and satellite at altitude $Z = 500$ km directly overhead and zenith angle $\theta = 0$, $10$, $20$ and $30$ degrees. Using the HV-5/7 model of atmospheric turbulence and an upper bound of $r_R^2 = 0.2$ for each atmospheric volume, we obtain 19 phase screens (Fig. \ref{fig:pubfig_atmos_multiplot}.a). Simulations of the downlink channel yield a beam profile which closely resembles the equivalent propagation of a Gaussian beam through vacuum, with loss in the channel primarily arising from medium-independent beam diffraction (Fig. \ref{fig:pubfig_atmos_multiplot}.b). In contrast, the uplink channel beam profile demonstrates the degrading effects of turbulence when applied to the early stages of the beam's propagation, with the profile showing a speckled pattern indicative of heavy beam wandering (Fig. \ref{fig:pubfig_atmos_multiplot}.c). Integrating the intensity of each beam's field over the receiver aperture, which we choose to be $a_r^\text{downlink} = 40$ cm and $a_r^\text{uplink} = 15$ cm, shows that the effective channel loss of the downlink channel is on the order of $\eta^\text{downlink} \sim 10^{-2}$, whereas the uplink channel has a much higher loss on the order of $\eta^\text{uplink} \sim 10^{-5}$.

Because of the stochastic nature of the atmospheric turbulence, we obtain the probability distribution $T^\text{atm}(\eta)$ for the uplink and downlink channels by simulating each 10 000 times (Fig. \ref{fig:pubfig_atmos_multiplot}.d). The downlink channel distribution is highly localised at $20$ dB, in agreement with other works as well as analytical models \cite{villasenor_alvarez_advanced_2023, pirandola_reply_2015, gonzalez-raya_satellite-based_2024}, since the dominant loss mode is the constant diffractive loss for finite receiver apertures. The distribution is skewed right in agreement with analytic atmospheric models using the Weibull distribution \cite{pirandola_satellite_2021}, where the sharp left side corresponds to perfect alignment and the right tail corresponds to long-term beam wandering effects. The uplink channel is substantially more random, commensurate with the small receiving aperture and highly random beam pattern, and shows channel losses ranging between $40$ and $70$ dB with a mean loss of approximately $50$ dB. This is in general agreement with the experimental data produced by \cite{ren_ground--satellite_2017}, which also records an average loss of $40-50$ dB for a ground-to-satellite channel of zenith height $500$ km and zenith angle $\theta = 0$ rad.

It should also be noted that the atmospheric channels may introduce a thermal-noise contribution to the signal arising from both fading noise as well as thermal input if the channel is operated during the day \cite{pirandola_satellite_2021, gonzalez-raya_satellite-based_2024}. \textcolor{black}{For ease, we assume night-time operation exclusively such that the effective mean photon number contributed by environmental noise is reduced to a negligible amount. (Ref. \cite{pirandola_satellite_2021} estimates $10^{-6}$ for typical downlink and $10^{-7}$ for typical uplink channels). This also justifies the appropriateness of the assumption of pure-loss channels for both the triple-satellite and ground-station-ground configuration \cite{pirandola_satellite_2021}. Furthermore, the assumption of night-time operation also justifies the use of the Hufnagel-Valley 5/7 model, which is empirically appropriate for atmospheric effects on a clear night \cite{stotts_improving_2023}.} In addition, we don't include here contributions from lesser atmospheric loss modes, such as backscattering or detector efficiency. These are relatively constant contributions to the loss on the order of $\sim 1$ dB and are generally negligible compared to the loss induced by beam broadening or wandering effects \cite{gonzalez-raya_satellite-based_2024, pirandola_satellite_2021}.

%

\end{document}